# Machine Learning-Assisted 3D Printing of Thermoelectric Materials of Ultrahigh Performances at Room Temperature

Kaidong Song,‡[a] Guoyue Xu,‡[a] A. N. M. Tanvir,‡[a] Ke Wang,[b] Md Omarsany Bappy,[a] Haijian Yang,[c] Wenjie Shang,[a] Le Zhou,[c] Alexander Dowling,[b] Tengei Luo *[a] and Yanliang Zhang*[a]

Thermoelectric energy conversion is an attractive technology for generating electricity from waste heat and using electricity for solid-state cooling. However, conventional manufacturing processes for thermoelectric devices are costly and limited to simple device geometries. This work reports an extrusion printing method to fabricate high-performance thermoelectric materials with complex 3D architectures. By integrating high-throughput experimentation and Bayesian optimization (BO), our approach significantly accelerates the simultaneous search for the optimal ink formulation and printing parameters that deliver high thermoelectric performances while maintaining desired shape fidelity. A Gaussian process regression (GPR)-based machine learning model is employed to expeditiously predict thermoelectric power factor as a function of ink formulation and printing parameters. The printed bismuth antimony telluride (BiSbTe)-based thermoelectric materials under the optimized conditions exhibit an ultrahigh room temperature $zT$ of 1.3, which is by far the highest in the printed thermoelectric materials. The machine learning-guided ink-based printing strategy can be highly generalizable to a wide range of functional materials and devices for broad technological applications.

## Introduction

Thermoelectric devices (TEDs) are solid-state energy converters that generate electricity when subjected to an external temperature gradient or create a temperature difference and act as solid-state coolers when provided with electric current. The ability of TEDs to convert heat into electricity and vice versa has sparked tremendous research interest in developing high-efficiency devices for waste heat recovery and solid-state cooling in the past two decades.[1–12] Two-thirds of the world's energy consumption remains dissipated as waste heat, and harnessing this wasted energy more efficiently can produce 15 terawatts of electrical power in the US alone.[13] Meanwhile, cooling and thermal management are essential to human comfort in buildings and vehicles, as well as to the reliable operation and longevity of electronic and medical devices. The solid-state nature of thermoelectrics makes it an attractive environmentally friendly technology for energy harvesting and cooling because it does not require moving parts or environmentally harmful refrigerants.[14]

The efficiency of thermoelectric materials is determined by the dimensionless figure of merit $zT = S^2\sigma\kappa^{-1}T$, where $S$ denotes the Seebeck coefficient, $\sigma$ is the electrical conductivity, $\kappa$ is the thermal conductivity, and $T$ is the absolute temperature.[15] Achieving high $zT$ requires improving the thermoelectric power factor $S^2\sigma$ while reducing the thermal conductivity.[15,16] Despite recent progress in increasing the $zT$ values, the reported high $zT$ materials still rely on conventional manufacturing methods, including hot pressing, arc melting, zone melting, and spark plasma sintering, which can only produce simple bulk structures at relatively high cost.[17–21] Moreover, the conventional methods require additional lengthy and costly fabrication processes to convert these bulk TE materials into useful devices. As a result, state-of-the-art commercial bulk TEDs still suffer from high performance and cost ratio,[22] which are not competitive enough compared with other energy conversion technologies. The lack of scalable and cost-effective manufacturing methods remains a long-standing challenge to produce high-performance TEDs with customizable shapes and form factors for end-use applications, which presents a major barrier to large-scale TED adoptions for energy harvesting and cooling.[23]

Three-dimensional (3D) printing technology has revolutionized manufacturing by creating intricate 3D structures from diverse materials, and it has recently been applied to thermoelectric fields.[24–39] A notable method in 3D printing is direct ink writing (DIW) or extrusion printing, which is widely used for printing concentrated viscoelastic inks into functional materials and devices.[33,40] Despite recent progress in printing thermoelectrics, printed thermoelectric materials still suffer from relatively low $zT$.[41] Meticulous tuning and optimization of thermoelectric ink formulation and printing parameters are required to achieve high thermoelectric performances while maintaining high printability and shape fidelity.

[a] Department of Aerospace and Mechanical Engineering, University of Notre Dame, Notre Dame, IN 46556, USA. E-mail: tluo@nd.edu; yzhang45@nd.edu
[b] Department of Chemical and Biomolecular Engineering, University of Notre Dame, Notre Dame, IN 46556, USA.
[c] Department of Mechanical Engineering, Marquette University, Milwaukee, WI 53233, USA.
‡ These authors contributed equally to this work

The optimization of thermoelectric ink formulation and printing parameters has traditionally relied on Edisonian methods such as one-variable-at-a-time experimental sensitivity analyses. These heuristic approaches require extensive expert knowledge and time and resource-intensive experimentation. The recent advancement of machine learning techniques presents unprecedented opportunities to accelerate the discovery of optimal material formulations and manufacturing processes, especially when facing high-dimensional problems with multiple input processing parameters and output properties of interest.[42–44] Machine learning methods such as Bayesian Optimization (BO) and Gaussian process regression (GPR)[42] have been successfully applied to optimize the sintering processes and the compositions of thermoelectric composites to achieve high thermoelectric power factors and zT.[30,31,45] These advancements highlight the role of machine learning in the thermoelectric field to enable more efficient to develop new thermoelectric materials and innovate their manufacturing processes.

This paper integrates extrusion printing of bismuth antimony telluride (BiSbTe) based thermoelectric inks with constrained BO and support vector machines (SVM) to discover the optimal ink formulation and printing parameters. An innovative water-based ink formulation is employed with a very small amount of Xanthan gum (X-gum) as a rheological modifier to adjust the ink viscosity and optimize viscoelastic behavior, which is crucial for producing intricate 3D structures during extrusion printing. An ultrahigh thermoelectric power factor of about 3000 $\mu Wm^{-1}K^{-2}$ and zT of 1.3 at room temperature is demonstrated by extrusion printing with these optimized inks, which is among the highest in the printed thermoelectric materials (Fig. 1C). In addition, intricate 3D structures are printed, demonstrating the potential to produce devices with complex and customizable shapes that are highly desired in practical applications where the heat source surfaces are often irregular.

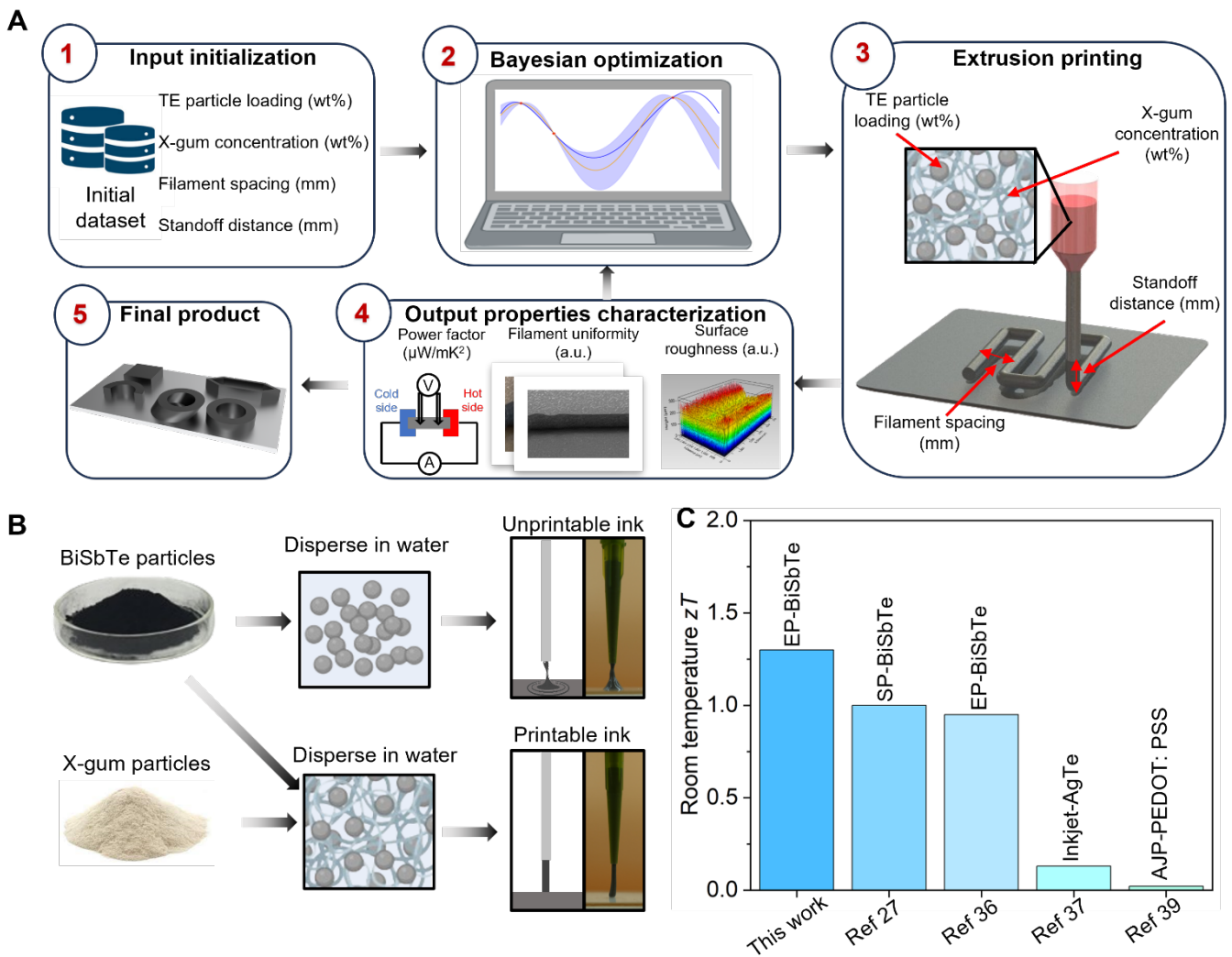

**Fig. 1** A) Workflow of the machine learning-assisted extrusion printing of thermoelectric inks, including the four input variables listed in box 1 and three out properties of interests in box 4. B) The printability of water-based thermoelectric inks with and without X-gum rheological modifier. C) Room-temperature thermoelectric figure of merit zT of our printed thermoelectric materials vs. best-reported values through printing in the literature.[29,38,39,41] (EP – extrusion printing, SP – screen printing, AJP – aerosol jet printing).

## Experimental

### Ink Preparation

It was established that incorporating a minor quantity of tellurium particles (Sigma-Aldrich, Burlington, MA) into the BiSbTe particle (Wuhan MCE Supply, Wuhan, China) mix could enhance the sintering process and improve the connectivity of the BiSbTe particles. During the sintering phase, these tellurium particles would melt at temperatures exceeding their melting point but still below that of BiSbTe. The thermoelectric powder was carefully weighed and blended with a specific solvent volume to prepare the ink. A rheological modifier, namely Xanthan gum (Sigma-Aldrich, Burlington, MA), was added to the DI water solvent. The amount of Xanthan gum added was proportionate to the overall mixture. The mixture underwent thorough mixing for homogenous ink consistency, first in a planetary centrifugal mixer for 30 minutes and then in a vortex mixer for an additional 10 minutes.

### 3D Printing and Sintering Process

The printing process was conducted using an in-house extrusion printer adapted from a commercial FDM machine. For the thermoelectric ink extrusion, we employed an 18-gauge nozzle with a 1.54 mm inner diameter (EFD Nordson, Vilters, Switzerland). The printer's reception bed was maintained at 40 °C to ensure printing quality. Printing was performed at a consistent tip travel speed of 2 mm s$^{-1}$.

A 51 μm thick HN-Kapton substrate was used as the base for printing the thermoelectric ink. Before printing, these Kapton films were precisely cut to size and thoroughly cleaned with methanol and isopropanol, aided by sonication. After printing, the samples were left undisturbed for 30 minutes to set. They were then subjected to a drying process at 200 °C for an hour in an inert atmosphere, which helped remove any residual solvent and rheological modifier. Post-drying, the samples were densified using a uni-axial hydraulic press, applying pressure up to 25 MPa. The final step involved sintering the samples at 450 °C for 90 minutes in a tube furnace under an inert atmosphere. Regarding the printability analysis, all printing paths were programmed using custom G-code scripts. For printing complex 3D structures, models were created using SolidWorks (Dassault Systems SolidWorks Corp, Waltham, MA) software and exported as STL files. These files were then processed using the Slic3r tools integrated into the control software of the FDM printer. G-code adjustments were made post-slicing, including setting the speed multiplier using MATLAB programming.

### Rheological Properties Characterization

The rheological properties of inks with and without Xanthan gum modifier were measured through a hybrid rheometer (HR-2 Discovery Hybrid Rheometer, TA Instruments, USA) with a 25 mm sandblasted ($Ra$ = 4.75 μm) parallel-plate measuring geometry and a 1 mm gap was utilized to perform all rheology measurements. Steady rate sweeps were conducted at a low strain (1%) for a shear rate range from 0.1 s$^{-1}$ to 100 s$^{-1}$ to detect the fluid viscosity and yield stress.

### Materials Characterization

Crystal structures of the synthesized structures were evaluated using Discover D8 XRD machine with a Cu Kα radiation with 1.54 Angstrom wavelength over a 2-theta range of 20–60. Microstructures and chemical compositions of the thermoelectric samples were examined using a scanning electron microscope (Helios G4 Ux Dual Beam) coupled with an energy-dispersive X-ray spectrometer (Bruker). A custom-built measurement setup following the Angstrom method is used for thermal diffusivity measurement.[30] Subsequently, the thermal conductivity is determined using the formula $\kappa = \alpha \rho C_p$, wherein $\alpha$ is the thermal diffusivity, $\rho$ is the density of the sample, $C_p$ is the constant pressure-specific heat capacity. The heat capacity was obtained from a previous publication.[29] The detailed implementation is described in Supporting Information 11.

### Post-Printing Processing

For the cold uni-axial pressing process, the printed and dried samples were densified using a hydraulic press, applying pressure up to 25 MPa for 10 mins before pressureless sintering.

For the hot isostatic pressing (HIP, AIP6-30H, American Isostatic Presses, Inc.) process, pressureless sintered samples were placed in a cylindrical molybdenum furnace under Ar atmosphere, with the thermocouples positioned proximate to both the samples and the molybdenum heating elements. The HIP temperature, pressure, and time were 480 ℃, 200 MPa, and 2 hours, respectively. The heating and cooling rates were set at 10 ℃/min and 7.5 ℃/min, respectively.

### Machine Learning and Optimization

In this work, the thermoelectric power factor is maximized while simultaneously ensuring good printability for 3D printing. As described in Fig. 1A, the manufacturing process contains four input controllable parameters (or decision variables), TE particle loading ($x_1$), X-gum concentrations ($x_2$), filament spacing ($x_3$), standoff distance ($x_4$), and three output variables, thermoelectric power factor ($y_1$), filament uniformity ($y_2$), and 3D structure surface roughness coefficient ($y_3$). Let the vectors $x_i$ and $y_i$ represent the inputs and outputs for experiment $i$. Here, we denoted the decision variables as $X = [x_1, ..., x_n] \in \mathbb{R}^{n \times 4}$, output variables as $Y = [y_1, ..., y_n] \in \mathbb{R}^{n \times 3}$, and formed the dataset as $\mathbb{D} = (X, Y)$.

GPR model is utilized to create a regression function $f(x)$ that maps the experimental conditions $x_i$ to the thermoelectric power factor $y_1$ while considering uncertainty (e.g., experimental variability). Two SVMs are introduced to learn manufacturing constraints $g(x_i)$ by classifying acceptable and unacceptable conditions. The acceptance threshold set for filament ($y_2$) is above 0.8, and for surface roughness ($y_3$) is below 0.05. Training data are assigned the labels 1 and -1 for acceptable and unacceptable experiments, respectively. Overall, the optimization problem can be formulated as:

$$\text{argmax}_x f(x)$$
$$\text{subject to } z_f(x) \geq 0, z_r(x) \geq 0$$

The SVMs $z_f(\cdot)$ and $z_r(\cdot)$ are integrated with BO as constraints. The Supporting Information 12 provides further details.

## Results and discussion

**Machine Learning-Assisted Optimization of Ink Formulation and Printing Parameters**

parameters (i.e., filament spacing and standoff distance). Thermoelectric power factor is chosen as the primary output property to be maximized, while the uniformity of the printed filament and the roughness of the printed structure are selected as the constraints that need to meet certain thresholds. In stage 2, we trained the GPR and SVM models and integrated them into constrained BO to determine the optimal ink formulation and printing parameters. GPR and SVM models are well-suited for small datasets (tens of datum). Moreover, GPRs explicitly model noisy observations, e.g., random experimental error, which is especially important for small datasets. The fact that GPR predicts uncertainty is also beneficial for its combination

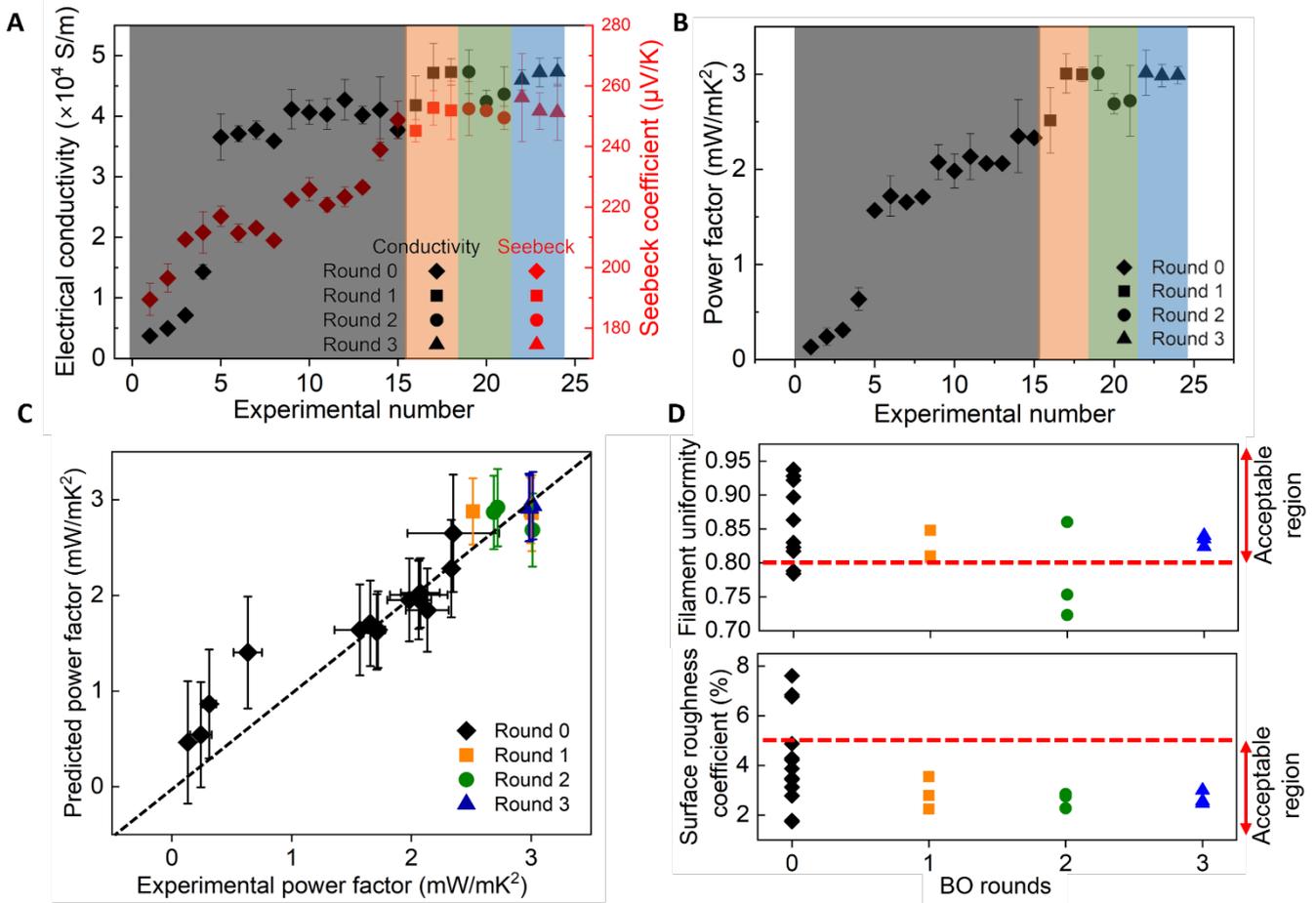

**Fig. 2** Room-temperature thermoelectric properties vs. experimental number (details of each experimental variable are summarized in Table S1). A) Electrical conductivity and the Seebeck coefficient. B) Power factor. C) The GPR parity plot illustrates the accuracy of our model predictions. Each round is denoted by a distinct color. Vertical error bars signify the model's predicted standard deviation, while horizontal error bars indicate the experimental standard deviation. D) The printability metrics for filament and surface roughness and their respective acceptable regions.

To accelerate the co-optimization of ink formulation and printing parameters, a method integrating high-throughput experimentation with constrained BO was developed. Fig. 1A shows a machine learning-assisted workflow organized into multiple stages. In stage 1, an initial input experimental data set was generated to train the machine learning model based on expert intuition. We identified four key input variables that significantly impact the outcomes of the printed materials, including two ink formulation parameters (i.e., TE particle loadings and X-gum concentrations) and two printing with BO. Other ML methods, e.g., deep learning networks, require orders of magnitude more data and do not explicitly consider experimental uncertainty. Stages 3 and 4 focus on printing samples based on the predictions from stage 2 and the characterization of their output properties (thermoelectric properties and shape accuracy). These new experimental data were fed back to the machine learning model to improve prediction accuracy further. The optimum ink formulation and printing parameters are identified after multiple iterations of stages 2 to 4 until there is no significant improvement in thermoelectric properties. Finally, in stage 5, the optimized ink formulation and printing parameters were employed to produce the highest-performing thermoelectric materials combined with high shape accuracy.

A significant challenge in extrusion printing thermoelectric materials is developing inks with high particle loading that exhibit suitable rheological properties and printability.[36] These properties are essential for achieving high thermoelectric performances while ensuring smooth printing and maintaining 3D structures with good geometric accuracy. While there is growing interest in using all-inorganic inks with inorganic binders for their viscoelastic properties,[32,33,46] the presence of a high concentration of organic solvents in these inks can introduce impurities and adversely affect the transport properties of the printed materials.[37] To address this issue, we formulated water-based inks for extrusion printing using X-gum as a rheological modifier to adjust the ink viscosity and optimize viscoelastic behavior. As depicted in Fig. 1B, the incorporation of X-gum transforms the ink's behavior from a low-viscosity liquid state to a printable medium with suitable viscosity for constructing complex 3D structures via extrusion printing, which significantly extends the ink's applicability and broadens the range of potential applications. Figs. S1 and S2 in the supporting information show detailed comparisons between the unmodified and X-gum-enhanced inks, and a comprehensive examination of their rheological behavior, including analyses of viscosity and shear modulus. The X-gum-enhanced thermoelectric inks exhibit shear-thinning and yield stress properties, which drastically enhance the capabilities of the aqueous thermoelectric inks for printing 3D structures and maintaining structural integrity.

Achieving desired thermoelectric properties and geometries in printed materials requires co-optimization of ink formulations and printing parameters. Ink formulation—particularly TE particle loading and X-gum concentration—and printing parameters like filament spacing and standoff distance crucially impact thermoelectric performance. Higher TE particle loading enhances structural density and thermoelectric properties by facilitating continuous pathways for charge carriers, thereby improving electrical conductivity and ZT, supported by previous studies.[32,33,36,37] However, excessive particle concentration increases viscosity, potentially disrupting uniform deposition and necessitating precise control of particle loading. Higher X-gum concentrations increase porosity (Fig. S13), influencing densification during sintering and affecting thermoelectric properties.[47,48] Tight filament spacing improves interface quality and reduces voids, enhancing structural density and performance, while excessive narrowing can lead to over-deposition. Similarly, standoff distance affects deposition accuracy. Larger distances produce discontinuous filaments, whereas smaller distances improve structural detail and reduce voids,[49] which is essential for high-performance devices. Optimizing these parameters via ML enhances the performance of thermoelectric devices and shape fidelity.

The Experimental Section and Supporting Information 3 detail the complete machine learning-assisted optimization process. The thermoelectric power factor (continuous variable) is treated as a primary objective to maximize, which is modeled using a GPR model. The uniformity of the printed filament and the roughness of the printed structures are treated as two constraints that must meet a certain threshold, which are described using the SVM classifier. The respective thresholds are set to be 0.8 for the filament uniformity and 0.05 for the surface roughness coefficient (surface roughness to filament diameter ratio) based on our observation of ink printability. Based on previous research, we adopted the thermoelectric material composition of $Bi_{0.4}Sb_{1.6}Te_3$ with 8 wt.% extra tellurium and the optimized sintering conditions of 90 minutes at 450 °C in a tube furnace with an inert gas environment. Our ink and printing optimization process involved testing 24 unique sets of decision variable values (ink formulations and printing parameters) detailed in Table S1 in the supporting information. The initial 15 data points were strategically chosen across a diverse range of input parameters related to ink formulation and printing parameters to effectively train the GPR model, enhancing its ability to detect key trends and interactions. An additional 9 data points were selected sequentially using BO to develop a probabilistic model that predicts performance and identifies optimal areas for improvement.

The machine learning-guided optimization leads to a notable thermoelectric property improvement at room temperature, as illustrated in Fig. 2A. The thermoelectric power factor shows appreciable increases, exceeding 3000 $\mu Wm^{-1}K^{-2}$ after four rounds of optimization (Fig. 2B). The parity plot depicted in Fig. 2C shows the GPR model's accuracy in predicting the thermoelectric power factor of the printed samples in each round. The error bars in the plot indicate the model's uncertainty and the inherent variability in experimental data. In addition, filament uniformity (> 0.8) and surface roughness coefficient (<0.05) of most experimental groups are within acceptable regions in the last three rounds (Fig. 2D). Figs. S6 and S7 in the supporting information elaborate on the complex interplay between different input and output parameters and the candidate's selections and model uncertainty during machine learning using heatmaps from sensitivity analyses.

**Characterization of the Printed Thermoelectric Materials**

The integration of high-throughput experimentation and constrained BO yields the optimal ink formulation and printing parameters: 83 wt.% particle loading, 0.5 wt.% X-gum using water solvent, 1.0 mm standoff distance, and 1.4 mm filament spacing. The temperature-dependent thermoelectric properties of the printed samples under the optimized conditions were measured in a temperature range of 20-200 °C. As shown in Fig 3A, the electrical conductivity exhibits a decreasing trend with increasing temperature, which is consistent with the behavior of highly doped BiSbTe-based materials.[50–52] The Seebeck coefficient shows a slight increase as the temperature rises, achieving a maximum of ~259 $\mu VK^{-1}$ in the 60 to 80 °C range. The printed sample shows a peak thermoelectric power factor of ~3000 $\mu Wm^{-1}K^{-2}$ at room temperature (Fig. 3B). The room-temperature thermal conductivity is measured to be 0.68 $W/mK$ using the Angstrom method. A room-temperature $zT$ of 1.3 is obtained for the printed samples under the optimized conditions, which is among the highest in printed thermoelectric materials.

mostly distributed along the grain boundaries (Fig. 3E). XRD patterns depicted in Fig. 3F indicate negligible variations among samples printed with and without rheological modifiers,

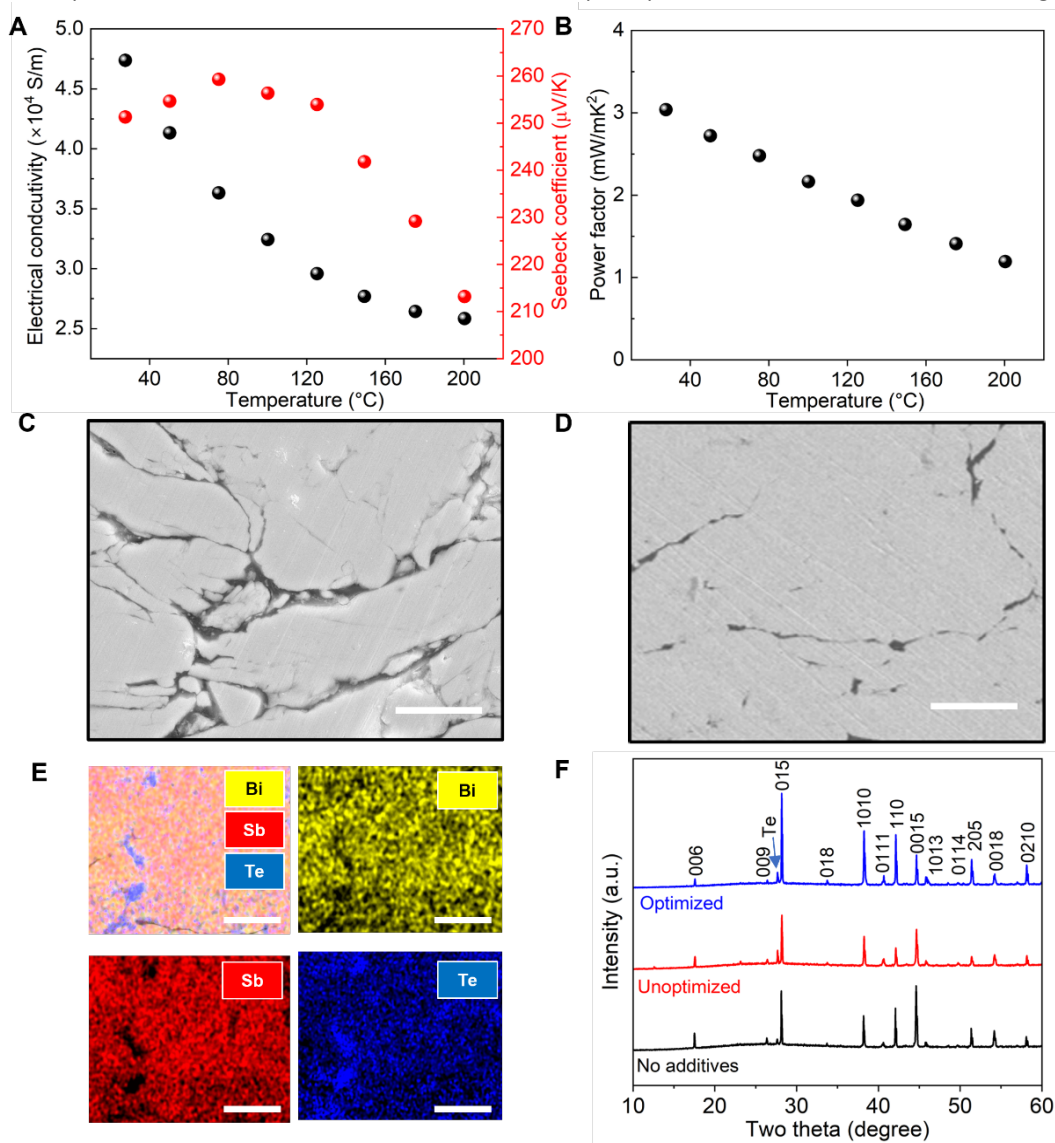

**Fig. 3** A-B) Temperature-dependent electrical conductivity, Seebeck coefficient, and power factor. C-D) Scanning electron microscope (SEM) images of the cross-section of samples printed using unoptimized and optimized inks (scale bar 10 μm). E) Elemental analysis of the sample printed using the optimized ink (scale bar 10 μm). F) XRD of printed samples with and without X-gum rheological modifier.

Scanning electron microscopy (SEM), energy-dispersive X-ray spectroscopy (EDS), and X-ray diffraction (XRD) were employed to characterize the printed materials under the unoptimized (62 wt.% particle loading and 4 wt.% X-gum) and the optimized (83 wt.% particle loading and 0.5 wt.% X-gum) conditions in order to understand the processing-structure-property correlations. Figs. 3C-D show SEM/EDS images of polished cross-sections of sintered samples printed from unoptimized and optimized inks. The unoptimized samples exhibited higher porosity (11.81%) compared with the optimized sample (5.43%), which was generated during the drying and sintering process due to the evaporation of the water solvent and X-gum rheological modifier compared to the optimized samples. Furthermore, EDS maps of optimized samples reveal that the excess tellurium is indicating $Bi_{0.4}Sb_{1.6}Te_3$ as the predominant phase with pristine tellurium as a secondary phase. This confirms that the drying and sintering process effectively evaporates the water solvent and eliminates the X-gum rheological modifier.

### 3D Printing of Optimized Thermoelectric Inks and Post-Printing Processing

We printed 3D structures using the optimized ink formulation and printing parameters to demonstrate the 3D printing capability. First, the printing performance of the thermoelectric inks was assessed by creating cubic structures ($8 \times 8 \times 8$ mm³), as shown in Fig. 4A. The optimized ink with the rheological modifiers yields relatively well-defined 3D cubic structures. One

Please do not adjust margins

of the key advantages of 3D printing is its ability to tailor the design of the printed structures, making it particularly suitable for producing thermoelectric elements that can match the surface of curved or irregular heat sources, such as exhaust pipes. Leveraging our optimized ink, we printed 3D structures of curved geometries, encompassing semi-circular and circular profiles, inclined tubes with a 60-degree angle, and hexagonal tubes (Fig. 4B). The successful printing of the inclined tubes indicates that our inks possess adequate yield stress to produce complex shapes for diverse applications.

Sintering plays a critical role in controlling the microstructures and properties of printed materials. Three sintering methods were investigated here: pressureless thermal sintering in a tube furnace (no press), cold uni-axial pressing followed by pressureless thermal sintering (cold press), and hot isostatic samples are similar to those of cold uni-axial pressing followed by pressureless sintering.

## Conclusions

In summary, the versatile ink-based printing method enables facile fabrication of thermoelectric materials of designed 3D shapes and competitive thermoelectric properties. The rapid printing combined with BO and GPR models significantly accelerates the discovery of optimized ink formulation and printing parameters in producing thermoelectric materials with enhanced thermoelectric performances. An ultrahigh thermoelectric power factor of 3000 $\mu Wm^{-1}K^{-2}$ and $zT$ of 1.3 at room temperature were achieved, which is significantly higher than the performance of previously reported 3D-printed pressing (HIP). The microstructure of the printed and sintered samples under no press, cold press, and HIP conditions have been shown in Figs. 3D and S9 in the supporting information. Comparative analysis of dimensional changes post-printing processing, as illustrated in Fig. S10 in the supporting information, reveals that HIP results in consistent shrinkage across all three dimensions, thereby preserving the intricate geometries of the printed structures. This is beneficial in applications involving irregular heat source surfaces. As shown in Figs. 4C and 4D, the thermoelectric properties of HIP-treated thermoelectrics. The ink-based printing can directly transform the starting thermoelectric particles into functional forms, which not only reduces material waste and manufacturing costs but also enables the fabrication of devices of desired shapes that can be seamlessly integrated with various heat sources. The machine learning-assisted ink-based printing framework is highly generalizable and can be used to manufacture a broad range of energy and electronic devices in a cost-effective and customizable manner.

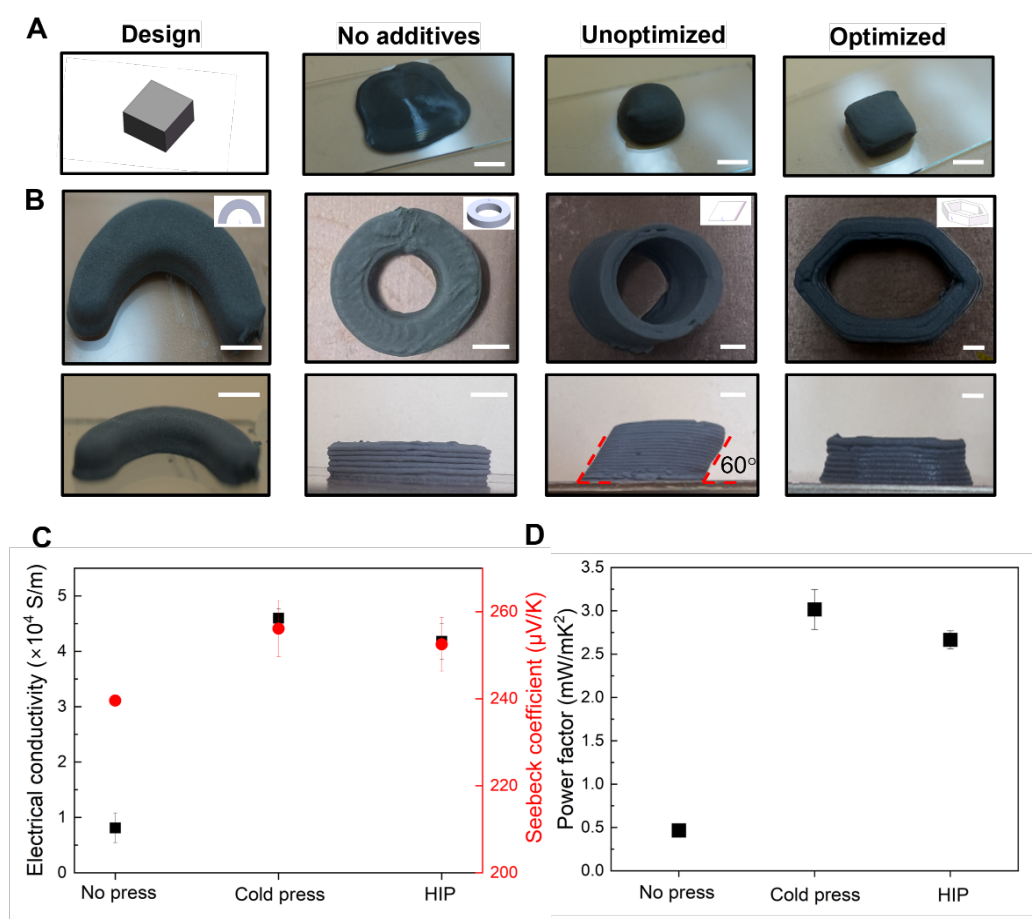

**Fig. 4** Printed 3D complex structures. A) Before and after optimization. (scale bar 4 mm) B) Complex 3D structures. (scale bar 4 mm) Comparison of samples C) electrical conductivity, the Seebeck coefficient, and D) power factor from corresponding post-printing processing methods.

## Author Contributions

K. S., G. X., T. L., and Y. Z. planned and designed the project. K. S. and A. N. M. T. synthesized the materials and conducted the experiments. K. S., G. X., K. W., and W. S. collected and analysed data. K. S., A. N. M. T., Md. O. B., and H. Y. characterized the samples. K. S., G. X., and A. N. M. T. wrote the manuscript. L. Z., A. D., T. L., and Y. Z directed the writing of the paper. A. D., T. L., and Y. Z supervised the project.

## Conflicts of interest

The authors declare no competing interests.

## Data availability

The data supporting this article have been included as part of the Supplementary Information.

## Acknowledgements

The authors would like to acknowledge support from the U.S. Department of Energy under award DE-EE0009103. Y. Z. would like to acknowledge funding support from the National Science Foundation under award CMMI-1747685 and the U.S. Department of Energy under award DE-NE0009138. K. W. also acknowledges ND Energy and the Patrick and Jana Eilers Graduate Student Fellowship for Energy-Related Research for providing financial support to advance this research.

## Footnotes

‡ These authors contributed equally to this work.